\documentclass[%
 aip,
 amsmath,amssymb,
 reprint,%
]{revtex4-1}

\usepackage{graphicx}
\usepackage{dcolumn}
\usepackage{bm}

\usepackage[utf8]{inputenc}
\usepackage[T1]{fontenc}
\usepackage{mathptmx}
\usepackage{etoolbox}
\usepackage{graphicx}
\usepackage{xcolor}
\usepackage{listings}
\usepackage{mhchem}

\definecolor{codeblue}{RGB}{0,70,140}
\definecolor{codegreen}{RGB}{0,120,0}
\definecolor{codered}{RGB}{160,40,40}
\definecolor{backgray}{RGB}{248,248,248}

\lstdefinestyle{scientific}{
    backgroundcolor=\color{backgray},
    basicstyle=\ttfamily\footnotesize,
    keywordstyle=\color{codeblue}\bfseries,
    commentstyle=\color{codegreen}\itshape,
    stringstyle=\color{codered},
    numberstyle=\tiny\color{gray},
    numbers=left,
    stepnumber=1,
    numbersep=8pt,
    showstringspaces=false,
    breaklines=true,
    frame=single,
    rulecolor=\color{gray!40},
    tabsize=4,
    captionpos=b
}

\makeatletter
\def\@email#1#2{%
 \endgroup
 \patchcmd{\titleblock@produce}
  {\frontmatter@RRAPformat}
  {\frontmatter@RRAPformat{\produce@RRAP{*#1\href{mailto:#2}{#2}}}\frontmatter@RRAPformat}
  {}{}
}%
\makeatother
\begin{document}

\preprint{AIP/123-QED}

\title{Fanpy 2.0: Wavefunctions Implementation and Analysis Tools for Flexible Ansatz Design}
\author{Krisztina Zsigmond}
\email{kzsigmond@ufl.edu}
\affiliation{Department of Chemistry and Quantum Theory Project, University of Florida, Gainesville, FL 32603, USA}

\author{Rugwed Lokhande}
\affiliation{Department of Chemistry and Quantum Theory Project, University of Florida, Gainesville, FL 32603, USA}

\author{Carlos E. V. de Moura}
\affiliation{Department of Chemistry and Quantum Theory Project, University of Florida, Gainesville, FL 32603, USA}

\author{Pratiksha B. Gaikwad}
\affiliation{Department of Chemistry and Quantum Theory Project, University of Florida, Gainesville, FL 32603, USA}

\author{Marco Mart{\'i}nez-Gonz{\'a}lez}%
\affiliation{Department of Chemistry \& Chemical Biology, McMaster University,\\ 1280 Main St.\ West, Hamilton, Ontario, L8S 4M1, Canada}

\author{Michelle Richer}%
\affiliation{Department of Mathematics and Statistics, University of Ottawa,\\ 75 Laurier Ave E, Ottawa, Ontario, K1N 6N5, Canada}

\author{Daniel F. Calero-Osorio}
\affiliation{Department of Chemistry \& Chemical Biology, McMaster University,\\ 1280 Main St.\ West, Hamilton, Ontario, L8S 4M1, Canada}

\author{Paul W. Ayers}
\email{ayers@mcmaster.ca}
\affiliation{Department of Chemistry and Chemical Biology, McMaster University, Hamilton Ontario L8S 4M1, Canada}

\author{Ram\'on Alain Miranda-Quintana}
\email{quintana@chem.ufl.edu}
\affiliation{Department of Chemistry and Quantum Theory Project, University of Florida, Gainesville, FL 32603, USA}

\date{\today}

\begin{abstract}

\texttt{Fanpy} is a Python library for developing new wavefunction methods. It enables users to quickly convert mathematical expressions into working code through a modular design based on the Flexible Ansatz for N-electron Configuration Interaction (FANCI) theory. This architecture facilitates a straightforward extension of the codebase. Here we present version 2.0 of the \texttt{Fanpy} package. This release includes several new wavefunction implementations, including coupled-cluster-inspired geminal approaches. A new analysis module enables a more detailed inspection of computational results and lays the groundwork for future features. In addition, the PySCF interface has been redesigned, and an interface to the PyCI package has been introduced to offload computationally expensive components. Finally, we introduce an improved software development environment, including automated testing and issue tracking.

\end{abstract}

\maketitle

\section{\label{sec:Introduction}The \texttt{Fanpy} Framework}

\texttt{Fanpy} is a free, open-source Python library for developing and testing multideterminant wavefunctions and related \emph{ab initio} methods in electronic structure theory.\cite{kim2023fanpy} Its design is motivated by a practical difficulty in the field: although many production-grade quantum chemistry packages support post-Hartree-Fock calculations, most are either closed-source\cite{g16, wernerMolproGeneralpurposeQuantum2012, qchem} or built around monolithic, performance-oriented codebases.\cite{psi4, GhentQChemPackage} Thus, implementing a new wavefunction ansatz requires deep familiarity with the entire package. \texttt{Fanpy} takes the opposite tradeoff. It deliberately compromises performance in favor of flexibility and ease of prototyping, with the explicit goal of letting researchers convert the mathematical formulation of a new method into a working implementation as quickly as possible. The flexibility of \texttt{Fanpy} is achieved by a highly modular design, in which wavefunctions, Hamiltonians, and optimization routines are implemented as independent, interchangeable components.

The library is built on the Flexible Ansatz for N-electron Configuration Interaction (FANCI) framework\cite{kim2021flexible}, which expresses any multideterminant wavefunction as a parameterized expansion in Slater determinants,

\begin{equation}
    \left| \Psi_{\mathrm{FANCI}} \right\rangle = \sum_{\mathbf{m} \in S_{\mathbf{m}}} f \left(\mathbf{m}, P\right) \left| \mathbf{m} \right\rangle,
    \label{eq:fanci}
\end{equation}

\noindent where $S_{\mathbf{m}}$ is the set of Slater determinants included in the wavefunction, $P$ is the set of wavefunction parameters, and the coefficient function $f$ is simply the overlap of the wavefunction with a determinant,

\begin{equation}
    f\!\left(\mathbf{m}, P\right) = \left\langle \mathbf{m} \middle| \Psi_{\mathrm{FANCI}} \right\rangle.
    \label{eq:overlap}
\end{equation}

This perspective is powerful because it places very different wavefunction families such as configuration interaction,\cite{shavitt1998history, popleVariationalConfigurationInteraction1977, boysElectronicWaveFunctionsFCI1950}
coupled cluster,\cite{bartlett2007coupled, cizekCorrelationProblemAtomic1966, evangelistaAlternativeSinglereferenceCoupled2011, evangelistaExactParameterizationFermionic2019, paldusCriticalAssessmentCoupled1999, shavittManyBodyMethodsChemistry2009,hendersonSenioritybasedCoupledCluster2014b,
steinSeniorityZeroPair2014,
boguslawskiLinearizedCoupledCluster2015},
matrix product states,\cite{mps2007, schollwockDensitymatrixRenormalizationGroup2011,woutersDensityMatrixRenormalization2014}
antisymmetrized product of geminals,\cite{tecmer2022geminal, hurleyMolecularOrbitalTheoryAPG1953, parrGeneralizedAntisymmetrizedProduct1956, parksTheorySeparatedElectron1958, allenELECTRONPAIRSBERYLLIUM1962, kutzelniggDirectDeterminationNatural1964, paldusDiagrammaticalMethodGeminals1972, paldusGeminalLocalizationSeparatedPair1971, shullNaturalSpinOrbital1959, Surjan1999, tecmerAssessingAccuracyNew2014a,johnsonSizeconsistentApproachStrongly2013,
limacherSimpleInexpensivePerturbative2014b,johnsonRichardsonGaudinMeanfield2020} 
and more, on the same mathematical footing: each is specified entirely by its choice of $P$ and $f$. As a direct consequence, implementing a new ansatz in \texttt{Fanpy} reduces, in the simplest case, to writing a function that evaluates Eq.~\ref{eq:overlap} for a given Slater determinant.

The Hamiltonian is handled in a complementary fashion, through its matrix elements in the determinant basis,

\begin{equation}
    \left\langle \mathbf{m} \middle| \hat{H} \middle| \mathbf{n} \right\rangle = \left\langle \mathbf{m} \middle| \sum_{ij} h_{ij} a_i^\dagger a_j + \tfrac{1}{2} \sum_{ijkl} g_{ijkl}, a_i^\dagger a_j^\dagger a_l a_k \middle| \mathbf{n} \right\rangle,
\end{equation}

\noindent so that the objective functions supported by \texttt{Fanpy}--- variational energy minimization,\cite{helgaker2013molecular, szabo2012modern, cramerEssentialsComputationalChemistry2013} and the projected Schr\"odinger equation,\cite{cullenGeneralizedValenceBond1996, johnsonStrategiesExtendingGeminalbased2017a, limacherNewWavefunctionHierarchy2016}---are all built from overlaps and Hamiltonian matrix elements alone.\cite{kim2023fanpy} The combination of wavefunction overlaps (Eq.~\ref{eq:overlap}) and Hamiltonian matrix elements, assembled in the objective function, is what makes \texttt{Fanpy}'s modular design possible: as long as a new wavefunction returns overlaps and a new Hamiltonian returns matrix elements, every other component of the package---objectives, solvers, optimizers, and analysis tools---can be reused without modification.

\begin{figure*}[t]
    \centering
    \includegraphics[width=\textwidth]{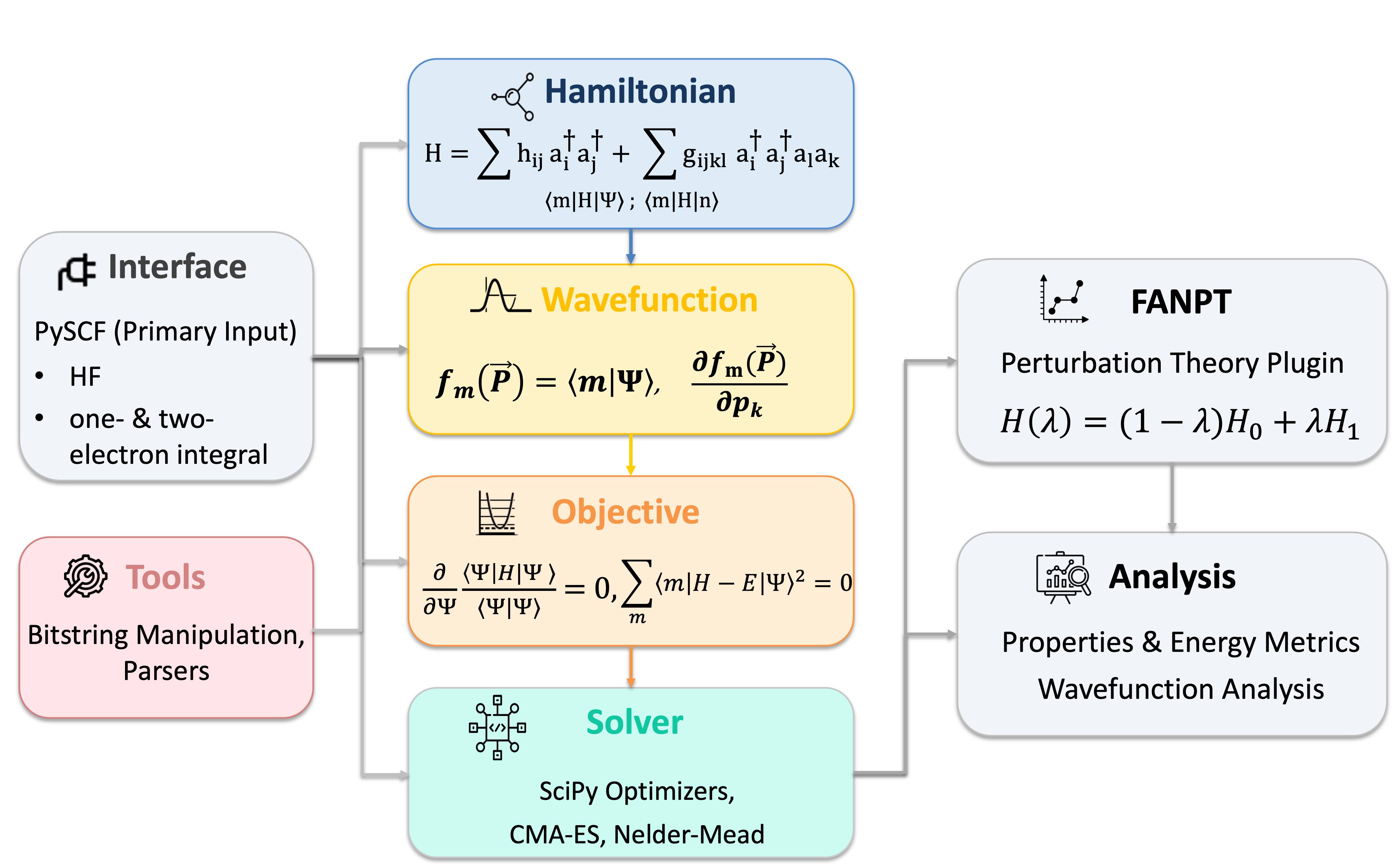}
    \caption{Modularity of \texttt{Fanpy} 2.0. The core computational pipeline flows through three coupled modules: the Hamiltonian, the wavefunction, and the objective (variational energy minimization or the projected Schr\"odinger equation), which is passed to one of the available solver methods. The interface module (PySCF) supplies Hartree-Fock references and one- and two-electron integrals, while utility tools provide shared bitstring manipulation routines. FANPT extends the core pipeline as a perturbative continuation framework for improving the solution of projected Schr\"odinger equations. An analysis module provides tools for computing properties such as natural orbitals and human-readable printing of wavefunction parameters to aid in the interpretation of results.}
    \label{fig: fanpy-modules}
\end{figure*}

Due to the modular nature of the FANCI framework, \texttt{Fanpy} is organized as a collection of independent components, each corresponding to one of the four objects needed to set up a post-Hartree-Fock calculation: a wavefunction, a Hamiltonian, an objective function representing the Schr\"odinger equation, and an algorithm for optimizing that objective function. A fifth unit provides shared utilities, most importantly, the tools for manipulating Slater determinants, that form the backbone of \texttt{Fanpy}. The resulting modular structure is, as summarized in Fig.~\ref{fig: fanpy-modules}, unchanged from version 1.0; the organization of the five modules is summarized below. A more detailed account of the base classes and their APIs is given in Ref.~\citenum{kim2023fanpy}.

\paragraph{Wavefunction module (\texttt{fanpy.wfn}).}
Each wavefunction in \texttt{Fanpy} is defined by its parameters and a method (\texttt{get\_overlap}) that returns its overlap with an arbitrary Slater determinant (Eq. \eqref{eq:overlap}). New wavefunctions can be introduced by subclassing the wavefunction base class and implementing \texttt{get\_overlap} and, when required, its parameter derivatives for gradient-based optimization. \texttt{Fanpy} 2.0 extends the wavefunction module with several new ansätze: coupled-cluster wavefunctions with arbitrary excitation spaces, including seniority-restricted and generalized variants; coupled-cluster-inspired geminal wavefunctions; and a neural-network wavefunction. These additions are described in Sec.~\ref{sec:new_wavefunctions}.

\paragraph{Hamiltonian class (\texttt{fanpy.ham}).}
Each Hamiltonian is defined by its representation in an orbital basis (i.e., one- and two-electron integrals) together with a function that returns its matrix element between two given Slater determinants. Restricted, unrestricted, and generalized electronic Hamiltonians are supported, along with the seniority-zero electronic Hamiltonian\cite{richardsonEigenstates$L0$Charge1967, calero-osorioSeniorityzeroWavefunctionParameterizations2025,bytautasSeniorityOrbitalSymmetry2011c} and the restricted Fock operator. Orbital optimization is available for any Hamiltonian that supplies derivatives with respect to orbital-rotation parameters.

\paragraph{Objective module (\texttt{fanpy.eqn}).}
The objective module combines a wavefunction and a Hamiltonian into the equation, or system of equations, to be solved. Three families of objectives are supported: variational minimization of the energy, the projected Schr\"odinger equation, and a local-energy expression suitable for variational quantum Monte Carlo sampling.\cite{kuritaVariationalMonteCarlo2015, neuscammanImprovedOptimizationCluster2016, neuscammanJastrowAntisymmetricGeminal2013, nightingaleQuantumMonteCarlo1998, sabzevariImprovedSpeedScaling2018, umrigarObservationsVariationalProjector2015} The module also provides fine-grained control over the optimization: parameters can be designated as active or frozen, checkpointing allows the optimization state to be saved and resumed, and nonlinear constraints can be imposed on the projected Schrödinger equation

\paragraph{Solver module (\texttt{fanpy.solver}).}
The solver module wraps the optimization algorithms used to minimize the energy or solve the projected Schr\"odinger equation. It exposes the relevant constrained and unconstrained optimizers from SciPy,\cite{2020SciPy-NMeth} brute-force eigenvalue decomposition for CI wavefunctions,\cite{BruteSciPyV1180} the CMA-ES evolutionary algorithm\cite{hansenPrincipledDesignContinuous2014} via \texttt{pycma},\cite{CMAESPycma2026} and Bayesian and tree-based derivative-free optimizers from \texttt{scikit-optimize}.\cite{headScikitoptimizeScikitoptimize2022} The modular interface makes it straightforward to add new domain-specific algorithms as the need arises.

\begin{figure}[h]
    \centering
    \includegraphics[width=0.85\linewidth]{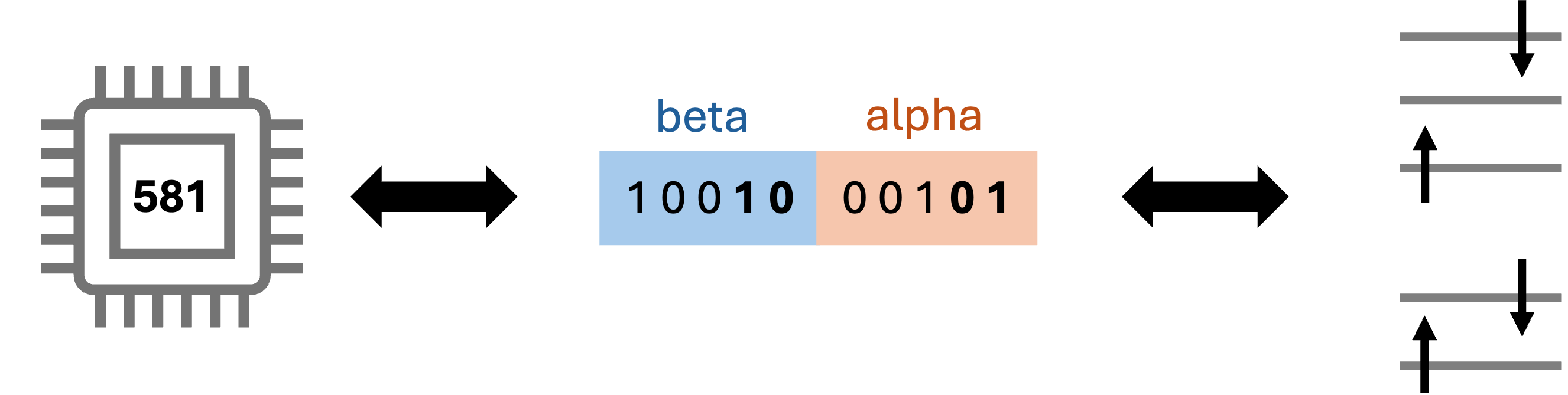}
    \caption{Bitstring representation in the FanPy package. The example Slater determinant is stored internally as the integer 581 (left). Its binary representation, 1001000101, is shown in the center. Bold digits indicate occupied orbitals, whereas non-bold digits indicate virtual orbitals. The first half of the bitstring (blue) contains the occupation numbers of the $\beta$ spin-orbitals, while the second half (orange) contains the occupation numbers of the $\alpha$ spin-orbitals. The corresponding orbital diagram is shown on the right.}
    \label{fig:bitstring_rep}
\end{figure}

\paragraph{Tool module (\texttt{fanpy.tools}).}
The tool module collects utilities that are used throughout the package, most importantly the routines for manipulating and enumerating Slater determinants (SDs), which serve as the common language of all other modules. SDs are represented as binary integers (with the positions of set bits indexing the occupied spin-orbitals). For example, the decimal number 581 is represented in binary as 1001000101 (see Fig.~\ref{fig:bitstring_rep}). The first half, 10010, encodes the occupations of the $\beta$ spin-orbitals, while the second half, 00101, encodes the occupations of the $\alpha$ spin-orbitals. This bitstring corresponds to the Slater determinant  
\begin{equation}
    a^{\dagger}_1 a^{\dagger}_3 a^{\dagger}_{\bar{2}} a^{\dagger}_{\bar{5}} |\rangle
\end{equation}
The \texttt{slater} module provides routines for converting, comparing, and operating on them, while the \texttt{sd\_list} generates determinants with specified excitation order, spin, or seniority and is used extensively when constructing projection spaces.

The modular theoretical formulation allows us to develop a new feature for one component and use it immediately with other parts of the code. This is what allows \texttt{Fanpy} to function as a prototyping environment rather than fixed implementation of any particular method. Furthermore, the code is easly extendable with new approaches, such as the \texttt{FANPT} module introduced in this article. 

The present article describes \texttt{Fanpy} 2.0, which extends the library along three main directions. First, a new \emph{interface} layer connects  \texttt{Fanpy} to two external packages PySCF\cite{sun2018pyscf} which now serves as the standard front end for Hartree-Fock calculations and integral generation, and PyCI\cite{richer2024pyci}, a CI library that can be used to offload the most expensive parts of a calculation while preserving \texttt{Fanpy}'s prototyping flexiblity. Second, \texttt{Fanpy} 2.0 introduces an implementation of the Flexible Ansatz for N-body Perturbation Theory (\emph{FANPT}),\cite{lokhande2024fanpt} a perturbative-continuation scheme that generates well-conditioned starting parameters for the projected Schr\"odinger equation, alleviating the initialization sensivity that often arises for strongly correlated systems and highly nonlinear wavefunctions. Third, the wavefunction module is expanded with four new families: coupled-cluster-inspired geminal wavefunctions that generalize the pair-coupled-cluster doubles formalism to broader excitation spaces,\cite{gaikwad2024coupled} seniority-restricted coupled cluster wavefunctions,\cite{lokhande2026srCC} an extended-hierarchy CI wavefunction,\cite{lokhande2026ehCI} and a neural-network wavefunction ansatz.\cite{kim2020systematic} All three directions are described in detail in the following sections. 

The source code, documentation, tests, and sample scripts for version 2.0 are available at \url{https://github.com/mqcomplab/Fanpy}.

\section{New Features}

Updates to the \texttt{Fanpy} codebase are motivated either by research projects (e.g., prototyping new wavefunctions) or by general software development, such as extending unit tests to cover the entire package. 

\subsection{Interface}

The guiding principle behind \texttt{Fanpy} is its flexibilty. In order to preserve it, we rely on external packages for standard quantum chemistry computations and faster backend operations. The interface submodule handles the communication between \texttt{Fanpy} and external packages. Currently, we utilize PySCF for Hartree-Fock calculations and PyCI can be used to improve the performance of \texttt{Fanpy}.

The PySCF interface extracts data \texttt{Fanpy} requires from HF calculations. This includes converting atomic to molecular-orbitals in addition to storing the number of electrons, number of spin-orbitals, energies, and related HF-derived quantities. Previously, \texttt{Fanpy} had wrapper functions for PySCF that extracted properties into a dictionary. However, this structure proved insufficient as we expanded the wrappers. For example, localizing the orbitals\cite{fosterCanonicalConfigurationalInteraction1960, marzariMaximallyLocalizedGeneralized1997, pipekFastIntrinsicLocalization1989} for embedding calculations required more properties from the PySCF calculation. 

We have refactored the wrapper into an interface class (\texttt{PYSCF}), where the constructor performs all transformations and sets up other necessary properties. This acts as the fundamental data structure for all subsequent calculations such as calculating the FCI matrix and performing the orbital localization mentioned above for embedding. Most importantly, we use \texttt{PYSCF} to set up electronic structure computations. Having a  uniform Hartree Fock datastructure allowed us to remove duplicate code in the old PySCF wrapper, improving code maintainability and stability.

Although the PySCF interface is required for all calculations unless the user has performed the HF calculations separately, the PyCI interface is optional. Utilizing it off-loads expensive determinant-space operations from \texttt{Fanpy}. This increases the performance, while maintaining flexibility. The interface is made up of two classes: a precomputation and an objective class. In the former, we set up the input for PyCI based on the \texttt{Fanpy} parameters and create an instance of the objective class. The latter performs all computations within PyCI, except for the computation of the overlap of the wavefunction with SDs. This allows us to implement a wavefunction in \texttt{Fanpy} and perform calculations for larger determinant spaces than a \texttt{Fanpy}-only set-up would allow.

\subsection{FANPT}

\texttt{Fanpy} 2.0 introduces an implementation of the Flexible Ansatz for N-body
Perturbation Theory (FANPT), a perturbative continuation framework designed
to improve the solution of the projected Schr\"odinger equation for general FANCI
wavefunctions.\cite{lokhande2024fanpt} In \texttt{Fanpy}, many wavefunctions are optimized by solving
a nonlinear system of projected equations rather than by direct diagonalization
or variational energy minimization. While this formulation is highly flexible,
the resulting nonlinear equations can be sensitive to the initial values of the
wavefunction parameters, particularly for strongly correlated systems or highly
nonlinear ans\"atze. FANPT provides a systematic way to generate better starting
points by propagating the parameters and energy from a simpler reference problem
toward the target molecular Hamiltonian.

The method is based on an adiabatic connection between an ideal Hamiltonian,
$\hat{H}_0$ (usually a Fock operator  $\hat{F}$, and the target Hamiltonian, $\hat{H}_1$,
\begin{equation}
\hat{H}(\lambda) = (1-\lambda)\hat{H}_0 + \lambda \hat{H}_1 ,
\end{equation}
where $\lambda$ is a coupling parameter. At $\lambda=0$, the projected
Schr\"odinger equation is solved for the simpler Hamiltonian. FANPT then advances
from $\lambda=0$ to $\lambda=1$ in a sequence of intermediate steps. At each
step, perturbative response equations are solved to estimate how the
wavefunction parameters and, when selected, the energy change with $\lambda$.
These propagated quantities are used as the initial guess for solving the
projected Schr\"odinger equation at the next value of $\lambda$. This process is
repeated until $\lambda=1$, where the Hamiltonian corresponds to the target
molecular Hamiltonian. In this way, FANPT acts as a continuation strategy that
connects an easier reference calculation to the final correlated calculation.
The overall workflow is shown in Fig 2.
\begin{figure}[htbp]
    \centering
    \includegraphics[scale=0.15]{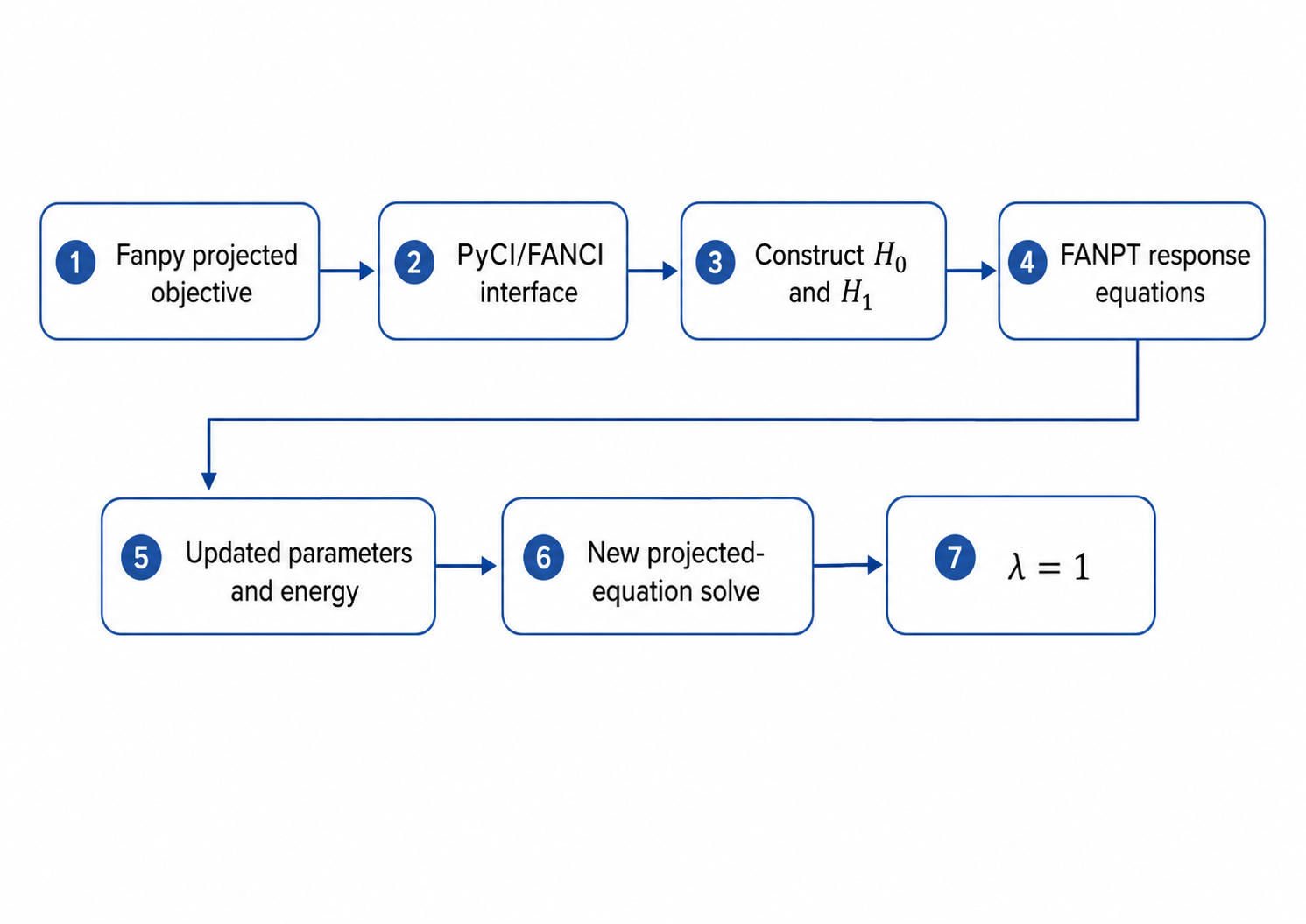}
    \caption{Schematic workflow of the FANPT implementation in \texttt{Fanpy} 2.0.}
    \label{fig:fanpt_workflow}
\end{figure}
The implementation is designed as a layer on top of the existing \texttt{Fanpy}
objective infrastructure. The user supplies a standard
\texttt{ProjectedSchrodinger} class from the \texttt{eqn} module, which already contains the wavefunction,
Hamiltonian, projection space, active parameters, and constraints needed for the
calculation. The high-level \texttt{FANPT} driver converts this objective to the
PyCI/FANCI interface, constructs the reference and target Hamiltonians, and
manages the $\lambda$-stepping procedure. This design keeps FANPT consistent with
the modular philosophy of \texttt{Fanpy}: the perturbative machinery does not introduce a
new wavefunction ansatz, but instead operates on existing \texttt{Fanpy} wavefunctions and
projected objectives.

The user-facing interface exposes the main choices that define a FANPT
calculation. These include the initial and final values of $\lambda$, the number
of intermediate steps, the final perturbative order, the quasi-approximation
order, the treatment of the energy, and optional resummation. Although a
complete FANPT calculation usually connects the reference and target
Hamiltonians by propagating from $\lambda=0$ to $\lambda=1$, the endpoints are
left adjustable so that users can restart calculations from an intermediate
value of $\lambda$, propagate only a selected segment of the path, or test the
behavior of the perturbative update over a shorter interval. A typical usage
pattern is shown below.

\begin{lstlisting}[language=Python]
from fanpy.fanpt import FANPT

# objective is a Fanpy ProjectedSchrodinger
# object containing the wavefunction, 
# Hamiltonian, projection space, active 
# parameters, and constraints.

fanpt = FANPT(
    objective,
    energy_nuc=energy_nuc,
    energy_active=True,
    final_order=2,
    lambda_i=0.0,
    lambda_f=1.0,
    steps=10,
    quasi_approximation_order=2,
)

fanpt_results = fanpt.optimize(
guess_params=wfn.params,
guess_energy=guess_energy,
)
\end{lstlisting}

In this example, the calculation begins from the ideal Hamiltonian and advances
to the target Hamiltonian in ten steps. At each step, FANPT solves the response
equations through second order and uses the resulting parameter update as the
initial guess for the next projected Schr\"odinger equation. The final output is a dictionary object containing the solution obtained along the perturbative path, including the optimized active variables and the final electronic energy. The optimized active variables are stored in \texttt{results["x"]}, while the final electronic energy is stored in \texttt{results["energy"]}.

Internally, the FANPT implementation separates the construction and solution of
the perturbative equations into several components. The base
\texttt{FANPTContainer} stores the quantities needed to build the linear FANPT
system, including the current parameters, $\lambda$, the ideal and
target Hamiltonians, the fluctuation potential, etc. This container-based design keeps
the low-level derivative and matrix construction separate from the high-level
workflow.

Two specialized containers are available depending on how the energy is treated.
In the energy-active formulation, the energy is included among the active
variables and is propagated together with the wavefunction parameters. In the
energy-free formulation, the energy is not treated as an active response
variable; instead, it is updated from the wavefunction after the parameter
correction has been applied. The high-level \texttt{energy\_active} option
selects the appropriate container automatically, allowing the user to switch
between these two workflows without changing the rest of the calculation setup.

At each perturbative order, FANPT solves a linear response equation of the form
\begin{equation}
\mathbf{J}(\lambda)\mathbf{x}^{(n)} = \mathbf{b}^{(n)},
\end{equation}
where $\mathbf{J}(\lambda)$ is the Jacobian of the projected FANCI equations, $\mathbf{x}^{(n)}$ is the $n$th-order response vector, and $\mathbf{b}^{(n)}$ is the constant-term vector generated by \texttt{FANPTConstantTerms}. This class collects the order-dependent terms that
enter the FANPT equations and uses the responses from lower orders when
constructing higher-order corrections. The solution and propagation steps are
handled by \texttt{FANPTUpdater}, which solves the response equations, updates
the active wavefunction parameters and energy, reconstructs the Hamiltonian at
the new value of $\lambda$, and refreshes the overlap information needed for the
next step.

FANPT is particularly useful as a convergence aid for flexible
wavefunction ans\"atze. Instead of requiring the user to manually construct a
sequence of guesses, \texttt{Fanpy} 2.0 can automatically propagate the wavefunction
parameters from a simpler Hamiltonian to the target Hamiltonian. As a result,
FANPT expands \texttt{Fanpy} from a platform for implementing new wavefunction forms into
a platform that also provides practical tools for solving the nonlinear equations
that those wavefunctions generate.

\subsection{Analysis}

An important step in quantum chemistry is the analysis of results after a calculation is done. \texttt{Fanpy} now has a new module with some initial analysis tools that will be expanded in future updates. For example, \texttt{Fanpy} can compute the natural orbitals\cite{lowdinQuantumTheoryManyParticle1955, davidsonNaturalOrbitals1972, helgaker2013molecular} after an electronic structure calculation. 

The analysis module also includes printers for wavefunction parameters. This alleviates the issue with examining them. \texttt{Fanpy} stores parameters as \texttt{numpy} arrays, thus printing out the array itself does not help with interpretability. The printer displays the data in a humanly readable way, allowing us to deduce which excitations, configurations, or geminals are important. For example, listing \ref{lst:analysis-printer-cc} shows the parameters of a CC wavefunction from the calculation in section \ref{sec:Examples}. In \texttt{Fanpy}, we represent the Slater Determinants as bitstrings. The excitation operator takes the electron from the orbital denoted with "-" and moves it to the orbital with the "+" sign. 

\begin{lstlisting}[label=lst:analysis-printer-cc, caption=Coupled Cluster amplitudes from the printer of the analysis module.]
Alpha    |  Beta     |  CC Parameter
-------------------------------------
111000+  |  11-0000  |   -0.49890337
111+000  |  11-0000  |   -0.41595214
11100+0  |  1-10000  |   -0.40618780
...
\end{lstlisting}

\subsection{New wavefunctions}\label{sec:new_wavefunctions}

The FANCI framework (Eq.~\eqref{eq:fanci}) provides a unified interface for developing and testing arbitrary multideterminant wavefunction ansätze: any wavefunction that exposes overlaps and their parameter derivatives through \texttt{get\_overlap} integrates immediately with the rest of the package. Building on this, \texttt{Fanpy} 2.0 introduces several new ansätze, described in the following subsections.

\subsubsection{Coupled-cluster inspired geminal wavefunctions}
In addition to the standard coupled-cluster (CC) models supported in version 1.0, the CC-wavefunction module (\texttt{fanpy/wfn/cc}) now supports a family of coupled-cluster-inspired geminal wavefunction\cite{gaikwad2024coupled}. These ans\"atze extend the pair-coupled-cluster doubles (pCCD) formalism to more general geminal excitation spaces.

The implementations build upon the one-reference orbital (1ro) formalism introduced by Limacher \emph{et al.}~\cite{limacher2013new}, which established the equivalence between the AP1roG geminal formalism and pCCD. Generalizing the 1ro construction to antisymmetric product of geminals (APG), antisymmetric product of geminals with disjoint orbital sets (APsetG),\cite{johnsonStrategiesExtendingGeminalbased2017a} and antisymmetric product of interacting geminals (APIG)\cite{johnsonStrategiesExtendingGeminalbased2017a, cassam-chenaiElectronicMeanField2010, cassam-chenaiElectronicMeanfieldConfiguration2006, rassolovGeminalModelChemistry2002, rassolovGeminalModelChemistry2004, rassolovGeminalModelChemistry2007, surjanStronglyOrthogonalGeminals2012, tecmer2022geminal,limacherNewWavefunctionHierarchy2016,
silverBilinearOrbitalExpansion1970,
silverNaturalOrbitalExpansion1969,
richerGraphicalApproachInterpreting2025,johnsonBivariationalPrincipleAntisymmetrized2022,
johnsonSingletGeminalWavefunctions2024} yields the doubles-only wavefunctions APG1roD, APsetG1roD, and APIG1roD. Each of these is further extended to a singles-and-doubles variant (APG1roSD, APsetG1roSD, APIG1roSD), in which seniority-breaking (increasing the seniority---the number of unpaired electrons, denoted $\Omega$---beyond the base seniority-zero ($\Omega=0$) in pCCD) single excitations are included to help recover dynamic correlation missing from the pair-only description and reduce models' sensitivity to orbital optimization.

All wavefunctions in this family are written as

\begin{equation}
\left| \Psi \right\rangle = \prod_{\mu \in \mathcal{E}} \left(1 + t_\mu \tilde{\tau}_\mu \right) \left| \Phi_0 \right\rangle,
\end{equation}

\noindent where $\left|\Phi_0\right\rangle$ is a seniority-zero reference determinant and $\mathcal{E}$ is an ansatz-specific pool of excitation operators. For the doubles-only generalized case (APG1roD), the pool consists of pair-annihilating, generalized double excitations,

\begin{equation}
\mathcal{E}_{\mathrm{APG1roD}} = \left\{\tau_{i\bar{i}}^{ab}\right\},
\end{equation}
which includes both pair-preserving ($b=\bar{a}$) and pair-breaking ($b\neq \bar{a}$) excitations, thereby extending beyond the pair-restricted excitation space of pCCD. The singles-and-doubles extension (APG1roSD) augments this pool with generalized single excitations,

\begin{equation}
\mathcal{E}_{\mathrm{APG1roSD}} = \left\{\tau_i^a, \tau_{i\bar{i}}^{ab}\right\}.
\end{equation}

The excitation pools for the APsetG- and APIG-
based variants follow the same construction, with the virtual creation indices restricted according to the underlying geminal ansatz; the corresponding equations and operator definitions are documented in the wavefunction classes.

The single-excitation manifold can be further constrained by seniority-breaking condition through the \texttt{s\_type} keyword inherited from the \textsc{PCCD} base class, which currently supports four options: \textit{sen-o} (singles allowed only out of doubly occupied orbitals), \textit{sen-v} (singles forbidden from forming virtual electron pairs), \textit{sen-ov} (both restrictions applied jointly), and \textit{free} (no restrictions). These conditions are enforced dynamically during overlap evaluation rather than at operator-pool construction. Combined with the three underlying 1roD/1roSD geminal families, the four \texttt{s\_type} options expose twelve distinct singles-and-doubles variants. For AP1roG-based wavefunctions, where singles are restricted to spin-conserving form, the corresponding \texttt{AP1roGSDSpin} variants use the pool $\mathcal{E} = \{\tau_i^a, \tau_{\bar{i}}^{\bar{a}}, \tau_{i\bar{i}}^{a\bar{a}}\}$, while the \texttt{AP1roGSDGeneralized} variants additionally permit spin-flip and spin-symmetry-breaking singles, $\mathcal{E} = \{\tau_i^a,\tau_i^{\bar{a}}, \tau_{\bar{i}}^a, \tau_{\bar{i}}^{\bar{a}}, \tau_{i\bar{i}}^{a\bar{a}}\}$ leading to sixteen newly supported wavefunction forms.

All of these wavefunctions are implemented as subclasses of \textsc{PCCD} in \texttt{fanpy/wfn/cc}, with each subclass defining only the excitation operator pool specific to the corresponding ansatz. New geminal variants can therefore be added with minimal implementation overhead while retaining full compatibility with the rest of \texttt{Fanpy}. Instantiation follows the same pattern as the other coupled-cluster wavefunctions: once the electron count, spin-orbital count, and \texttt{s\_type} keyword (default: \textit{sen-o}) are provided, the wavefunction object constructs the allowed excitation pool from the reference determinant according to the selected ansatz. For the doubles-only variants the \texttt{s\_type} keyword has no effect, since no single excitations are present. The resulting object plugs directly into the Hamiltonian, objective, and solver modules; an example calculation is provided in Sec.~\ref{sec:Examples}.

\subsubsection{Seniority-restricted coupled cluster wavefunctions}

Fanpy 2.0 includes new seniority-restricted coupled cluster (sr-CC)
wavefunctions.\cite{lokhande2026srCC} Recall that the seniority ``quantum number'', $\Omega$, counts the number of unpaired electrons in
a determinant relative to a chosen pairing of spatial orbitals. This provides a
physically useful way to organize electronic correlation: pair-preserving
excitations remain in the seniority-zero sector, whereas broken-pair excitations
introduce higher-seniority components.\cite{calero-osorioSeniorityzeroWavefunctionParameterizations2025,bytautasSeniorityOrbitalSymmetry2011c,doi:10.1021/acs.jctc.5c01892} By filtering the cluster operator
according to both excitation rank and seniority, sr-CC provides a flexible way
to interpolate between compact pair-based methods and more complete
excitation-based coupled-cluster expansions.

The implemented wavefunctions retain the exponential coupled-cluster form
\begin{equation}
    |\Psi_{\mathrm{sr\text{-}CC}}\rangle = e^{\hat{T}}|\Phi_0\rangle ,
\end{equation}
but restrict selected components of $\hat{T}$. \texttt{Fanpy} 2.0 introduces three sr-CC wavefunctions: \texttt{CCSDsen0}, \texttt{CCSDTQsen0}, and \texttt{CCSDsen2TQsen0}. The first two restrict the doubles and quadruples to seniority 0 excitations, respectively. The last wavefunction restricts doubles to seniority two and quadruples to seniority 0. 
\begin{equation}
    \begin{split}
        |\Psi_{\mathrm{sr\text{-}CCSD(0)}}\rangle &= e^{\hat{T}_1 + \hat{T}^{\Omega=0}_2} |\Phi_0\rangle \\ 
         |\Psi_{\mathrm{sr\text{-}CCSDTQ(0)}}\rangle &= e^{\hat{T}_1 + \hat{T}_2 + \hat{T}_3 + \hat{T}^{\Omega=0}_4} |\Phi_0\rangle \\
         |\Psi_{\mathrm{sr\text{-}CCSDT(2)Q(0)}}\rangle &= e^{\hat{T}_1 + \hat{T}_2 + \hat{T}^{\Omega=2}_3 + \hat{T}^{\Omega=0}_4} |\Phi_0\rangle ,
    \end{split}
\end{equation}

From a software perspective, these wavefunctions inherit from the coupled-cluster class, and limit the allowed excitation ranks and the seniority-filtered excitation operators on top of it. The parent coupled-cluster infrastructure handles parameter indexing, overlap evaluation,
derivative evaluation, caching, and compatibility with the objective and solver
modules. Thus, the sr-CC implementation illustrates how \texttt{Fanpy} can be used to
rapidly prototype physically motivated wavefunction truncations without
redesigning the surrounding optimization framework.

Due to inheritance, sr-CC wavefunctions follow the same initialization pattern as other
coupled-cluster wavefunctions in \texttt{Fanpy}. Once the number of electrons and spin
orbitals are specified, the wavefunction object constructs the allowed
excitation operators from the reference determinant according to the seniority
restrictions of the selected ansatz. The resulting object can then be passed to
the same Hamiltonian, objective, and solver modules used for other \texttt{Fanpy}
wavefunctions.

\begin{lstlisting}[language=Python]
from fanpy.wfn.cc.ccsd_sen0 import CCSDsen0
from fanpy.eqn.energy_oneside import EnergyOneSideProjection
from fanpy.solver.equation import minimize

nelec = 8
nspin = 16

# sr-CCSD(0): singles + seniority-zero doubles
wfn = CCSDsen0(nelec, nspin)

\end{lstlisting}

This example illustrates how the sr-CC implementation fits naturally within the general \texttt{Fanpy} design. The seniority restrictions are encoded in the sr-CC wavefunction classes themselves, so the Hamiltonian, objective, and solver components can be used without sr-CC-specific modifications. In other words, users can exchange one sr-CC ansatz for another
without changing the Hamiltonian, objective, or solver interface. For example,
the wavefunction definition above can be replaced by
\texttt{CCSDTQsen0(nelec, nspin)} or
\texttt{CCSDTsen2Qsen0(nelec, nspin)} to access the higher-rank
seniority-restricted models.

\subsubsection{Extended-hierarchy CI wavefunctions}

While sr-CC restricts excitation operators based on seniority, the extended-hierarchy configuration interaction (ehCI) takes a different approach.\cite{lokhande2026ehCI, kossoskiHierarchyConfigurationInteraction2022}  It selects determinants based on a scheme that organizes the CI expansion using both excitation rank and seniority: 
\begin{equation}
    h = \alpha_1 e + \alpha_2 s ,
\end{equation}
where $e$ is the excitation degree, $s$ is the seniority number, and
$\alpha_1$ and $\alpha_2$ define the ordering of the excitation--seniority
lattice. The implemented patterns, which determine $\alpha_1$ and $\alpha_2$, include positive slope diagonals, negative
 slope diagonals, horizontal chess horse, and vertical chess horse.

The main software role of this class is to automate the construction of
structured CI spaces. For a selected hierarchy pattern and cutoff, the
\texttt{hCI} class first determines the allowed $(e,s)$ sectors. It then uses
\texttt{Fanpy}'s existing \texttt{sd\_list} functionality to generate only those Slater
determinants that match the requested excitation order and seniority. Thus,
ehCI reuses \texttt{Fanpy}'s determinant-generation tools rather than introducing a
separate determinant-selection backend.

\begin{lstlisting}[language=Python,basicstyle=\ttfamily\scriptsize]
from fanpy.wfn.ci.hci import hCI

nelec = 8
nspin = 16

# Original hCI/PSD ordering
wfn = hCI(
    nelec,
    nspin,
    hci_pattern="pos_diag",
    hierarchy=2.0,
)

# Alternative ehCI ordering
wfn = hCI(
    nelec,
    nspin,
    hci_pattern="neg_diag",
    hierarchy=2.0,
)
\end{lstlisting}

After the determinants are generated, duplicates are removed and the selected
space is passed to the standard \texttt{Fanpy} CI class. As a result, ehCI adds
new excitation--seniority filtering logic while reusing the existing \texttt{Fanpy}
machinery for determinant handling, coefficient storage, overlap evaluation,
Hamiltonian integration, objectives, and solvers.

\subsubsection{Neural network Wavefunction ansatz}
The neural-network (NN) wavefunction ansatz, originally proposed by Kim,\cite{kim2020systematic} represents another addition to \texttt{Fanpy} 2.0's wavefunction module. Neural networks are universal approximators and can, in principle, represent any functional form, including the overlap $\langle \mathbf{m} | \Psi \rangle$ of a wavefunction with a Slater determinant.\cite{barronUniversalApproximationBounds1993, cybenkoApproximationSuperpositionsSigmoidal1989, funahashiApproximateRealizationContinuous1989, hornikMultilayerFeedforwardNetworks1989} The key idea is to employ a neural network as a nonlinear parameterization of the CI coefficients: the wavefunction takes a CI-like form,

\begin{equation}
    \left| \Psi_{\mathrm{NN}} \right\rangle = \sum_{\mathbf{m}} f(\mathbf{m})\, \left| \mathbf{m} \right\rangle,
\end{equation}

\noindent where $\mathbf{m} \in \{0,1\}^{2K}$ is the occupation-number representation of the Slater determinant $\left|\mathbf{m}\right\rangle$ (with $K$ spatial orbitals) and the coefficient $f(\mathbf{m})$ is the scalar output of a feedforward network parameterized by per-layer weight matrices $\{\mathbf{W}^{(\ell)}\}_{\ell=0}^{L}$ and activation functions $\{\sigma^{(\ell)}\}$. 

Two network architectures for $f(\mathbf{m})$ are implemented in the \texttt{network} submodule of the core \texttt{wfn} module. The first is a feedforward network with a multiplicative readout: the network applies the standard sequence of linear transformations and nonlinear activations through $L-1$ hidden layers, and the
$N$-dimensional pre-activation of the output layer follows element-wise nonlinear activation and is subsequently collapsed to a scalar by a product over its components,

\begin{equation}\label{eq: nn-multiplicative-readout}
    f(\mathbf{m}) = \sigma^{(L)}\!\left( \prod_{i=1}^{N} z_i^{(L)} \right),
    \qquad
    z_i^{(L)} = \sum_{j} W_{ij}^{(L)}\, h_j^{(L-1)},
\end{equation}
with $h^{(\ell)}$ denoting the hidden activations at layer $\ell$. The multiplicative structure ensures size consistency of the wavefunction and admits a natural initialization: setting the weight matrices to appropriately scaled identity matrices recovers the Hartree-Fock ground state exactly, providing a stable and physically motivated starting point for the optimization. The second architecture is a standard single-output feedforward network, in which the final layer produces a scalar directly, offering a simpler alternative for cases where the multiplicative readout is not required.\cite{coeMachineLearningConfiguration2018,
deyMachineLearningQuantum2023,
ghoshConfigurationInteractionTrained2021,
ranoEfficientMachineLearning2023}

Both ans\"atze are implemented in the \texttt{fanpy/wfn/network} module. Continued development of neural-network-based wavefunction ans\"atze in \texttt{Fanpy} will be hosted in the same location.  

\subsection{Miscellaneous developments}

In addition to the new features explained above, \texttt{Fanpy} 2.0 introduces "quality of life" improvements and other maintenance related updates. These include an automated test pipeline, utilizing GitHub features to facilitate code development, and updating \texttt{Fanpy} to work with \texttt{numpy 2.0}.\cite{numpy_v2.0}

Unit tests are a crucial, though often overlooked, task of software development. An extensive test suite ensures quality control and  gives developers confidence extending and improving the code base. \texttt{Fanpy} uses \textit{pytest}\cite{PytestDocumentation} in combination with GitHub Actions\cite{GitHubActions2026} to run checks automatically when changes are proposed or when the code changes on the \texttt{main} branch. This allows us to test \texttt{Fanpy} in a clean environment with multiple Python versions and catch dependency updates.  

We have also created a holistic software development environment. For this, we make use of the following GitHub features: branch protection, issues, and projects. First, we have established branch protection rules to safe guard the \texttt{main} branch, i.e. where the production ready code lives. Currently, a code review is required to merge changes, and the aforementioned testing suite is run automatically. This has greatly enhanced the quality of the code and prevented multiple bugs. Second, we rely on issues to track software development tasks. For this, we have created issue templates for bug reports, and feature requests. Especially for the former, it is crucial we get the required information to debug the issue. Lastly,  issues are automatically added to a GitHub project, which gives us an overview of the stage a given issue is in, similar to a task management apps.

\section{\label{sec:Examples}Examples}

\texttt{Fanpy}'s modular structure allows users to prototype new wavefunction ansätze, experiment with different Hamiltonians and objectives. We refer the reader to Ref.~\citenum{kim2023fanpy} for a description of how to implement these classes. 

The steps for running electronic structure calculations are as follows: 

\begin{enumerate}
    \item Perform a Hartree Fock calculation
    \item Set up the wavefunction and the Hamiltonian
    \item Initialize the objective 
    \item Optimize the Schrodinger equation
    \item Analyze results 
\end{enumerate}

\noindent Note that due to the flexibility of \texttt{Fanpy}, we can easily change the methods at any given step, while the rest of the steps remain the same. Here we show an example calculation for the \ce{BeH2} molecule with the AP1roGSD seniority-o wavefunction. 

In the first step, we perform the HF calculation with PySCF and use the \texttt{Fanpy}'s interface to extract the data from the results:

\begin{lstlisting}[language=Python]
from pyscf import gto, scf
import fanpy.interface as interface

print('# PySCF calculation for alpha = 2.75...')
mol = gto.M(atom = [['Be', ['0.0', '0.0', '0.0']], ['H', ['0.0', '-1.275', '2.75']], ['H', ['0.0', '1.275', '2.75']]],
            unit = 'B',
            basis = 'sto-6g')

myhf = scf.HF(mol)
myhf.kernel()
new_guess, _, stable, _ = myhf.stability(return_status=True)

if not stable:
    print("HF stability failed, trying Newton method...")
    myhf = myhf.newton().run(new_guess, myhf.mo_occ)
    _, _, stable, _ =  myhf.stability(return_status=True)

if not stable:
    raise RuntimeError("HF stability failed")
# PySCF interface
pyscf_interface = interface.pyscf.PYSCF(myhf)
\end{lstlisting}

\noindent The user can also provide one- and two-electron integrals from other software packages in the form of \texttt{numpy} arrays.\cite{kimGBasisPythonLibrary2024,
sunLibcintEfficientGeneral2015,chuikoModelHamiltonianPythonscriptableLibrary2024} 

Next, we define the wavefunction and the Hamiltonian:

\begin{lstlisting}[language=Python]
import numpy as np
from fanpy.wfn.cc.ap1rog_generalized_NEW import AP1roGSDGeneralized

wfn = AP1roGSDGeneralized(pyscf_interface.nelec, pyscf_interface.nspinorb)
wfn.assign_params(wfn.params + 0.1 *  (np.random.rand(*wfn.params.shape) - 0.5))
from fanpy.ham.restricted_chemical import RestrictedMolecularHamiltonian

ham = RestrictedMolecularHamiltonian(pyscf_interface.one_int, pyscf_interface.two_int, update_prev_params=True)
\end{lstlisting}

\noindent Here, the wavefunction and Hamiltonian can be swapped out to any child class of \texttt{BaseWavefunction} and \texttt{BaseHamiltonian}. 

After this, we connect the Hamiltonian and wavefunction instances in an objective. Here we use the projected Schr\"odinger equation for which we have to define the projection space: 

\begin{lstlisting}[language=Python]
from fanpy.eqn.projected import ProjectedSchrodinger
from fanpy.tools.sd_list import sd_list

# Projection space
exc_orders = [1, 2, 3, 4, 5, 6]
pspace = sd_list(pyscf_interface.nelec, pyscf_interface.nspinorb, num_limit=None, exc_orders=exc_orders, spin=0)

objective = ProjectedSchrodinger(wfn, ham, energy_type="compute", pspace=pspace)

\end{lstlisting}

\noindent The \texttt{"compute"} energy type means that we will determine the energy based on the wavefunction parameters, and not treat it as another parameter to optimize. 

Finally, we optimize the objective with one of the solvers:

\begin{lstlisting}[language=Python]
from fanpy.solver.system import least_squares

results = least_squares(objective, xtol = 10**(-4), gtol=10**(-4), ftol=10**(-4))

\end{lstlisting}

\noindent Again, we have prioritized flexibility for optimizing the objective. For example, we can use the wrappers to solvers such as scipy, use the PyCI interface to speed up the calculation, or optimize the wavefunction parameters with FANPT. 

Optionally, we can examine the results with the tools from the analysis module. For example, we can showcase the wavefunction parameters with:

\begin{lstlisting}[language=Python]
from fanpy.analysis.printer.cc import print_excitation_operators_as_determinants

print_excitation_operators_as_determinants(wfn)
\end{lstlisting}

\noindent which produces the output in Listing \ref{lst:analysis-printer-cc}

\section{Conclusion and future development}

The updates presented here demonstrate the flexibility and extensibility of \texttt{Fanpy}. Many of the newly implemented wavefunction methods build upon existing components, such as the coupled-cluster class, which reduced the development effort required for their implementation. At the same time, the Hamiltonian and solver modules remained unchanged, illustrating the modularity of the codebase. The newly introduced analysis module further enables detailed examination of computational results.

Beyond new wavefunction implementations, version 2.0 expands the interoperability of \texttt{Fanpy}. The interface to PyCI allows computationally demanding parts of calculations to be offloaded, while the PySCF interface streamlines Hartree--Fock calculations and their integration into \texttt{Fanpy} workflows.

Future development of \texttt{Fanpy} will focus on improving the performance of the existing codebase, extending the capabilities of the analysis module, further integration into the QC-Devs ecosystem,\cite{chanTaleHORTONLessons2024} and enhancing the documentation. These efforts aim to further support the rapid development and exploration of novel wavefunction methods.

\begin{acknowledgments}

We acknowledge the support from the National Science Foundation CAREER award CHE-2439867 (RAMQ), the Canada Research Chairs (CRC-2022-00196; PWA), NSERC (RGPIN/06707-2024, ALLRP/592521-2023; PWA), and the Digital Research Alliance of Canada.

\end{acknowledgments}

\section*{Data Availability Statement}

\texttt{Fanpy} is free and open-source software, licensed under the GNU GPL version 3 or later.
Its source code is available at \href{https://github.com/mqcomplab/Fanpy}{https://github.com/mqcomplab/Fanpy}.

\section*{Notes and references}

\bibliography{aipsamp} 

\end{document}